\documentclass[%
 aip,
 amsmath,amssymb,
reprint,%
]{revtex4-1}

\usepackage{graphicx}
\usepackage{dcolumn}
\usepackage{bm}
\usepackage{amssymb}
\usepackage[utf8]{inputenc}
\usepackage[T1]{fontenc}
\usepackage{mathptmx}
\usepackage[english]{babel}
\usepackage[version-1-compatibility,
per-mode=reciprocal,
emulate=units,
decimalsymbol=.,
loctolang={DE:ngerman,UK:english},
separate-uncertainty=true,
expproduct=cdot,
range-phrase={ -- },
range-units=single]{siunitx}

\DeclareSIUnit[number-unit-product = \,]{\promille}{\textperthousand}
\DeclareSIUnit[number-unit-product = \,]{\ppm}{ppm}
\DeclareSIUnit[number-unit-product = \,]{\Kkb}{\kelvin \cdot \textit{k}_\mathrm{B}}
\DeclareSIUnit[number-unit-product = \,]{\joule}{\text{J}}
\DeclareSIUnit[number-unit-product = \,]{\joulepermolkelvin}{\joule\per\mol\per\kelvin}

\DeclareSIUnit{\sqrthz}{\ensuremath{\sqrt{\text{\hertz}}}}
\DeclareSIUnit{\phinoise}{\Phi_{\rm 0}\per\sqrthz}
\DeclareSIUnit{\phinought}{\Phi_{\rm 0}}

\begin{document}

\preprint{AIP/123-QED}

\title[Development of a novel calorimetry setup based on metallic paramagnetic temperature sensors]{Development of a novel calorimetry setup based on \\metallic paramagnetic temperature sensors}

\author{Andreas Reifenberger}
 \email{Andreas.Reifenberger@kip.uni-heidelberg.de}
\author{Andreas Reiser}
\author{Sebastian Kempf}
\author{Andreas Fleischmann}
\author{Christian Enss}
\affiliation{Kirchhoff Institute for Physics, Heidelberg University, Im Neuenheimer Feld 227, 69120 Heidelberg, Germany}

\date{\today}

\begin{abstract}
We have developed a new micro-fabricated platform for the measurement of the specific heat of low heat capacity mg-sized metallic samples, such as superconductors, down to temperatures of as low as $10\,\mathrm{mK}$. It addresses challenging aspects of setups of this kind such as the thermal contact between sample and platform, the thermometer resolution, and an addenda heat capacity exceeding that of the samples of interest (typically $\mbox{nJ/K at }20\,\mbox{mK}$). The setup allows us to use the relaxation method, where the thermal relaxation following a well defined heat pulse is monitored to extract the specific heat. The sample platform ($5 \times 5\, \mathrm{mm^2}$) includes a micro-structured paramagnetic \underline{Ag}:Er temperature sensor, which is read out by a dc-SQUID via a superconducting flux transformer. In this way, a relative temperature precision of $30\,\mathrm{nK/\sqrt{Hz}}$ can be reached, while the addenda heat capacity falls well below $0.5\,\mathrm{nJ/K}$ for $T < 300\,\mathrm{mK}$. A gold-coated mounting area ($4.4 \times 3 \, \mathrm{mm^2}$) is included to improve the thermal contact between sample and platform.
\end{abstract}

\maketitle

\section{\label{sec:01_Introduction}Introduction}
Determining the specific heat of samples down to very low temperatures $T$ provides detailed insight into low-energy excitations like electronic, phononic, or magnetic subsystems of solids. The specific heat $c_p$, linked directly to the entropy $S$ of a system via $\Delta S = \int (c_p/T) \mathrm{d}T$, is a valuable tool to study such degrees of freedom as well as phase 
transitions\cite{Phillips1960,Galazka1980,Pobell2007,Esquinazi1998,Barron1999,JohnsonGuangyonggxu}. Studying phenomena dominated by quantum effects often require measurements at ultra-low temperatures, which are nowadays rather easily accessible by means of commercially  available dilution refrigerators \cite{Lounasmaa1979,Uhlig2002a}. Measurement devices to determine the specific heat of samples down to $50\,\mathrm{mK}$ are commercially available\footnote{See for example PPMS Dilution Refrigerator Option by Quantum Design, Inc., 10307 Pacific Center Court, San Diego, CA 92121, USA}, however systems for measurements down to $10\,\mathrm{mK}$ are still non-standard.

There are essentially three experimental approaches to determine the specific heat of a sample at low temperatures\cite{Stewart1983}. In the first approach, the adiabatic measurement, the sample is thermally decoupled from its environment and a temperature increase $\Delta T$ due to a small and well-known heat input $Q$ is measured. The heat capacity can then simply be calculated from its definition
\begin{equation}C = \lim\limits_{\Delta T \rightarrow 0}\frac{Q}{\Delta T} \enskip .\end{equation}
At very low temperatures and sample sizes well below \SI{1}{\gram}, it becomes experimentally more and more challenging to realize the adiabatic boundary conditions while allowing the sample to thermalize before the actual measurement. \\
The second approach known as ac-calorimetry overcomes these limitations by explicitly allowing for a finite thermal contact between sample and environment, thus enabling measurements even under high pressure. A sinusoidal heat input is applied and the resulting sinusoidal change of sample temperature, which is inversely proportional to the heat capacity of the sample, is recorded \cite{Sullivan1968,Wilhelm2003,Ga2019}. Changes in the heat capacity $\Delta C$, especially for small samples below \SI{1}{\milli\gram}, can be monitored precisely, while the absolute value of $C$ is typically less accurate in this method. \\
The third approach, the so-called relaxation method, relies on solving the heat flow equations for a given heat input into the sample and does not require adiabatic conditions, either\cite{Bachmann1972}. 
To extract a sample's heat capacity, it is placed onto a platform (addenda), which provides a heater and a thermometer. An adequate thermal coupling between sample and addenda is obtained by using an adhesive such as vacuum grease. Monitoring the time-resolved thermal relaxation $\Delta T(t)$ of the system following a well defined heat input allows the determination of the specific heat at a given bath temperature $T$ by a numerical fit. Typically, the sample is heated by less than 3\,\% of the bath temperature during such a heat pulse. The (weak) thermal coupling between addenda and thermal bath defines the relaxation time scale, typically some seconds or minutes. The limited thermal conductivity of glass-like vacuum grease at very low temperatures results in the well known $\tau_{\mathrm{2}}$-effect. To a certain extent, this can be handled by adjusting the numerical model accordingly \cite{Hwang1997}. \\
The calorimeter presented in this paper is of the latter kind, as it is the most suitable for the measurement of absolute values of $C$ for our samples of interest - mainly mg-sized metallic samples at temperatures of as low as \SI{10}{\milli\kelvin}. In addition, this approach allows us to bring in our expertise in producing highly precise, ultra-fast metallic magnetic thermometers\cite{Fleischmann2005,Kempf2018}, which turns out to be crucial for devices of this kind.

\section{\label{sec:02_Challanges}Challenges}
With the measurement scheme introduced above in mind, we can identify the main challenges that one faces at cryogenic temperatures when designing a heat capacity measurement device:

\textbf{Thermometry:} The temperature resolution $\delta T$ must be good enough for resolving the thermal response of a heat pulse both in temperature ($\delta T / T \lesssim 0.03\,\%$) and time (typically $\Delta t \approx 20\,\mathrm{ms}$ between two measurement points) to allow for a meaningful fit, i.e. keep systematic errors due to a limited resolution in temperature and time negligible. The time resolution is constrained by the necessity to resolve the $\tau_{\mathrm{2}}$-effect, which is crucial to keep systematic fitting errors negligible, and can vary depending on the used adhesive thickness and the sample contact area. Furthermore, the readout power of the temperature sensor must typically be far less than $0.1\,\mathrm{pW}$ at lowest temperatures to minimize the effects of parasitic heating. A standard resistance thermometer such as RuOx read out by a lock-in amplifier with a typical noise level in the order of \SI{2}{\nano\volt\per\sqrt\hertz} does not meet these criteria -- limiting its use to temperatures above $50\,\mathrm{mK}$ (due to the required time resolution). 

\begin{figure}
\includegraphics[width=250pt]{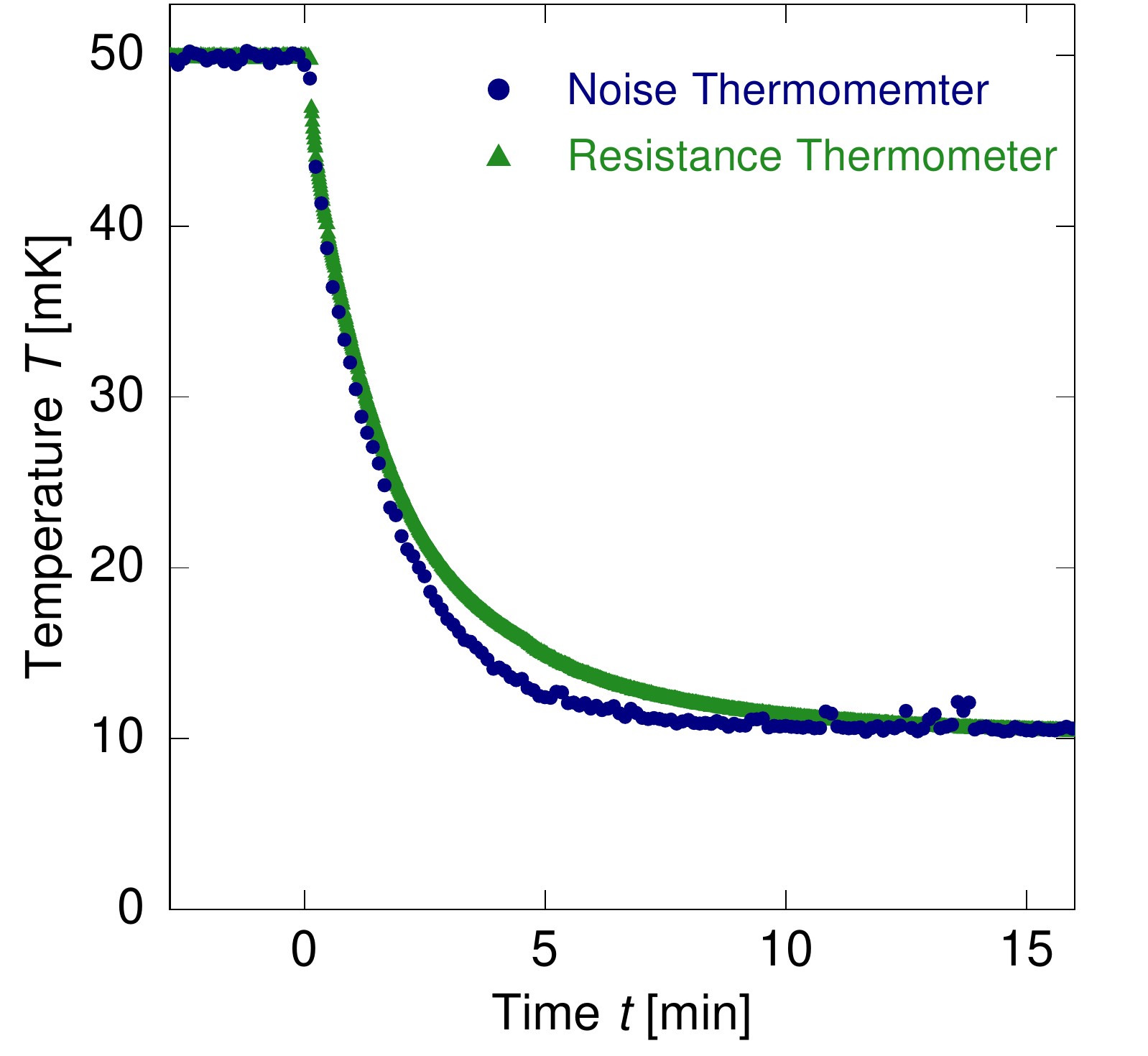}
\caption{\label{fig:fig01_thermometry}Time-resolved temperature response of a noise thermometer (blue dots) compared with a resistance thermometer (green triangles) after a step-like decrease of the bath temperature at time $t\,=\,0$. Data courtesy of Hempel et al.\cite{Hempel2017}}
\end{figure}

We would like to stress that even an excellent readout scheme would not be sufficient to overcome these limitations. Next to the limitations set by the intrinsic electron-phonon coupling of resistance thermometers\cite{Wellstood1994}, a significant nuclear heat capacity leading to slow internal relaxation times is another drawback of (most) resistance thermometers. Figure~\ref{fig:fig01_thermometry} displays the recorded temperature of a RuOx-type resistance thermometer and a current sensing noise thermometer \cite{Fleischmann2019}, both firmly attached to the mixing chamber of a dilution refrigerator. At $t=0\,\mathrm{min}$ the heat input at the mixing chamber is decreased by a factor of about \num{25}, allowing the temperature to fall from $50\,\mathrm{mK}$ down to $10\,\mathrm{mK}$. We can clearly see a time delay of several minutes for the temperature measured via the resistance thermometer in comparison to the noise thermometer. The reason for this rather slow intrinsic thermalization behavior is likely caused by nuclei carrying electric quadrupole moments (such as $^{101}$Ru). These interact with the local crystal field giving rise to an energy splitting and hence creating an additional Schottky-like heat capacity contribution. Due to the intrinsically weak coupling to the phonon bath at lowest temperatures, we observe a delayed fall in temperature. The noise thermometer, on the other hand, is based on quadrupole free bulk silver where this effect is not expected. 

This discussion underlines that resistance thermometers are not a suitable choice for building a calorimeter operating at temperatures down to \SI{10}{\milli\kelvin}.


\vspace{0.3cm}
\textbf{Addenda Heat Capacity:}
With our focus on superconducting, mg-sized samples, the heat  capacity $C_{\rm s}$ of such a sample can become rather small in the mK temperature range. Our goal is to resolve sample heat capacities of some nJ/K. Since we can only determine the total heat capacity \begin{equation} C_{\rm tot} = C_{\rm s} + C_{\rm add} + C_{\rm vg} \quad ,\end{equation}
we need to obtain $C_{\rm s}$ by subtracting the contribution of the addenda $C_{\rm add}$ and of the vacuum grease $C_{\rm vg}$. Therefore, a very small addenda heat capacity, well below $1\,\mathrm{nJ/K}$, is essential. This was achieved by a micro-structured addenda platform, allowing for minimal material use. Furthermore, superconducting materials, Nb and Al, were used wherever possible.

\vspace{0.3cm}
\textbf{Thermal Contact}
Another important aspect when designing a heat capacity experiment for temperatures down to $10\,\mathrm{mK}$ is the thermal contact between sample and platform. Due to the glassy structure of the vacuum grease (reducing its thermal conductivity by some orders of magnitude) and due to the thermal boundary resistance between sample, vacuum grease, and addenda, one observes a limited thermal conductance $\kappa_{\rm 2}$ between sample and addenda platform. Hence, at temperatures below about $30\,\mathrm{mK}$, $\kappa_{\rm 2}$ gets comparable to the thermal conductance $\kappa_{\rm 1}$ between the addenda platform and the thermal bath. This means that we heat up the addenda significantly, while the sample itself is only warmed up marginally. As a consequence, it is unfeasible to keep the $\tau_{\rm 2}$-effect small enough at those temperatures. However, this is crucial when performing the numerical fitting in order to not suffer from systematic errors. In this paper, we discuss the approach of an electronic contact between conductive samples and the platform to overcome this limitation.

\section{\label{sec:03_Experimentals}Experimental Setup}
\begin{figure}
	\includegraphics[width=250pt]{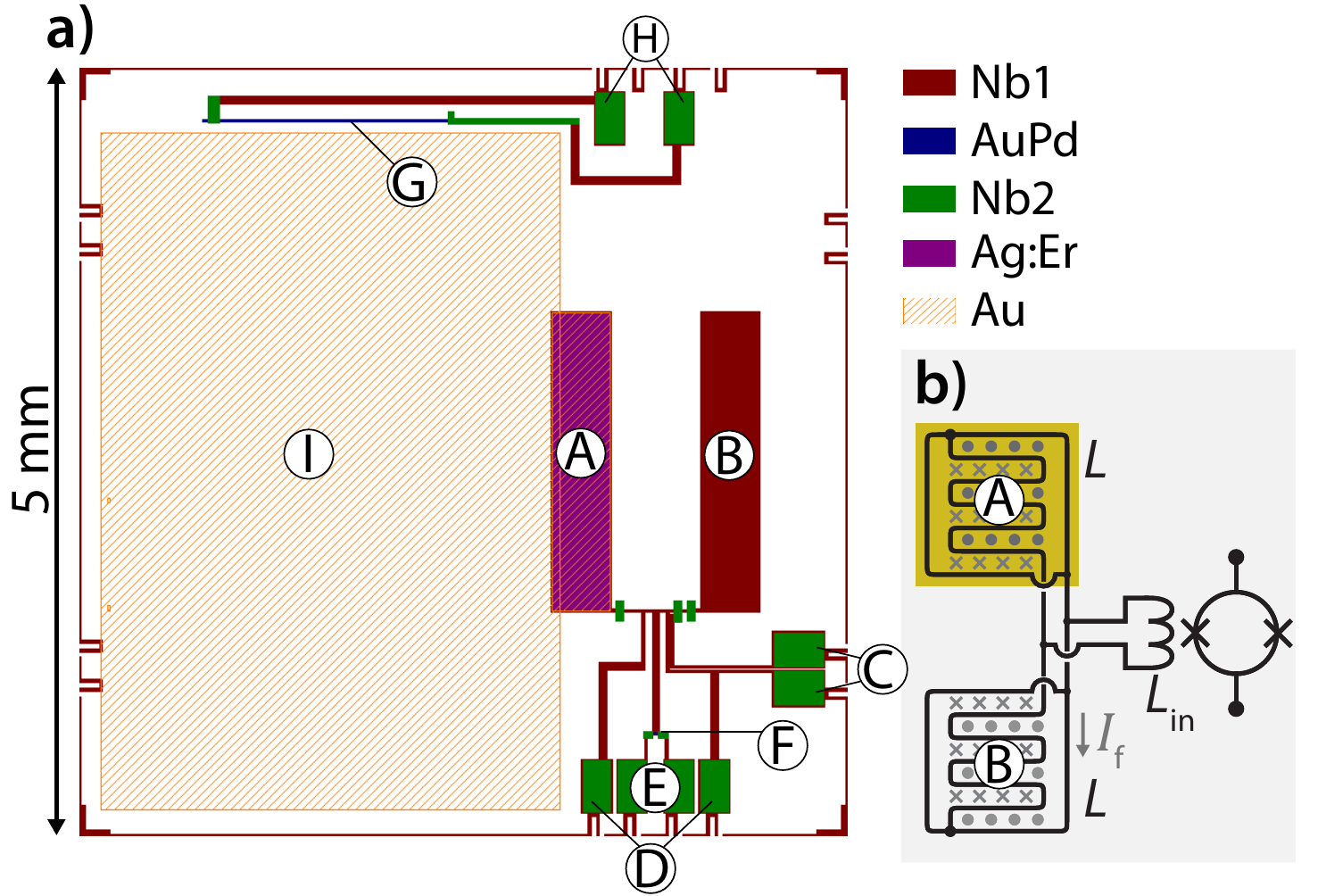}
	\caption{\label{fig:fig02_HeatCapChip}(a) Schematic drawing of the layout of the developed heat capacity chip. The materials deposited onto the chip are indicated by colors (Nb1 in red, AuPd in blue, Nb2 in green, \underline{Ag}:Er in purple, and Au in orange). Nb1 and Nb2 indicate sputtered Nb films deposited at different steps during fabrication. Further details are discussed in the text. (b) The equivalent circuit diagram of the two superconducting coils (A) and (B) connected to the input coil of a current sensing dc-SQUID. Coil (A) is covered by the sensor material \underline{Ag}:Er and a layer of Au (orange box).}
\end{figure}

We developed a novel micro-fabricated calorimeter read out by a two-stage dc-SQUID setup based on an idea proposed earlier \cite{Reifenberger2014}. A schematic drawing of the chip, $5\,\mathrm{mm} \times 5\,\mathrm{mm}$ in size, is shown in figure~\ref{fig:fig02_HeatCapChip}\,(a).

\begin{figure}[h!]
	\includegraphics[width=250pt]{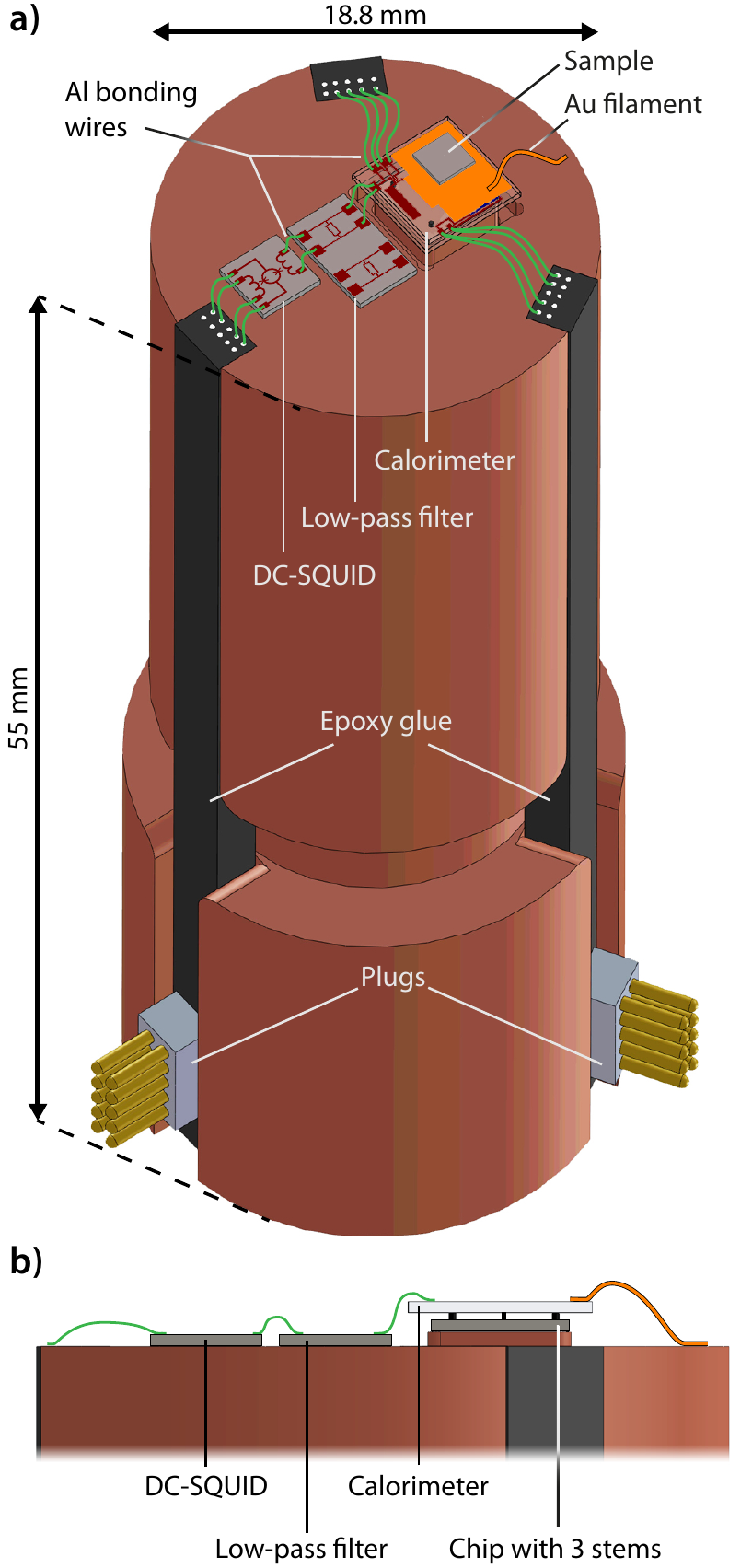}
	\caption{\label{fig:fig03_Tsingy}Schematic drawing of the full experimental setup. (a) Isometric view: The front face of a copper rod hosts the calorimeter chip, the low-pass filter and the dc-SQUID. Bonding wires are drawn in green (Al) and orange (Au). The SQUID array employed to operate the two-stage SQUID setup is hosted on a separate copper holder not shown here. Further details are provided in the text. (b) The side view of the upper-most part of the experimental setup illustrates the placement of another chip with three stems to support the calorimeter, as described in the text. }
\end{figure}
The chip is made from a thermally oxidized, high purity silicon substrate and hosts two superconducting meander-shaped pickup-coils (A) and (B) made of niobium connected in parallel. The inductance of each is $L = \SI{60}{\nano\henry}$ and was chosen such that we can maximize the output signal when using current sensing dc-SQUIDs with an input inductance of about \SI{20}{\nano\henry} for readout. There is a \SI{250}{\nano\meter} thick insulating layer of sputtered SiO$_{2}$ on top of each pickup coil (not shown for clarity).  A persistent current $I_{\rm f}$ can be injected into the circuit at temperatures below $T_{\rm c}$ by driving a current $I_{\rm0}$ through the pickup coils (bondpads (D)). A small fraction of the loop near (F) is then driven normal-conducting by applying a heat pulse ($\approx \SI{1}{\milli\second}$, via bondpads (E)) through a  \SI{25}{\ohm}-resistor made of AuPd (F) allowing flux to enter the circuit. This persistent current generates a magnetic field $B_{\rm 0} \propto I_{\rm f}$, which orients the electronic spins in the paramagnetic material \underline{Ag}:Er\footnote{The notation \underline{Ag}:Er indicates a diluted Ag host material doped with some $100\,\mathrm{ppm}$ of magnetic Er atoms} placed on top of one of the pickup coils (A). A temperature change of the chip will change the magnetization $M(T)$ of the \SI{1.6}{\micro\meter} thick \underline{Ag}:Er film with an erbium concentration of $x_{\rm Er} = \SI{350\pm50}{ppm}$. This in turn creates a screening current through the superconducting meanders. A dc-SQUID, whose input coil $L_{\rm in}$ is connected in parallel to the two meanders via bondpads (C) by superconducting Al bonding wires, can detect this screening current. Figure~\ref{fig:fig02_HeatCapChip}\,(b) shows the corresponding equivalent circuit diagram. By means of a two-stage SQUID setup\cite{Clarke2004} operated in flux-locked loop mode, a voltage $U \propto \Delta M$ can be monitored, which gives information about the temperature of the paramagnetic temperature sensor, while keeping the power dissipation of the dc-SQUID near the calorimeter chip itself at a minimum. This detection scheme is applied in metallic magnetic calorimeters (MMCs) for single particle detection and has been discussed in great detail elsewhere\cite{Fleischmann2005}.

Once the persistent current is prepared and the corresponding magnetization curve has been calibrated against the bath temperature of the cryostat, we can perform an actual heat capacity measurement. For this, we drive a heating current $I_{\rm h}$ through a temperature-independent AuPd-resistor (G) with $\mathrm{RRR} \approx 1,35$ in order to generate well-defined joule heating. Typical heating times during operation, depending on sample heat capacity and thermal contact, vary in the range of seconds to some minutes. With the help of two independent readout schemes (4-wire measurement across the AuPd-resistor and/or a current meter in series to the galvanically decoupled current source), we estimate the accuracy in $Q$ to be better than \SI{1}{\percent}.

This heat is absorbed by the chip itself as well as the sample of interest mounted on the electroplated gold layer (I). Bonding wires made of Al and Au, which couple the chip weakly to the thermal bath, allow for the appropriate thermal relaxation. By monitoring the full temperature response of such a sequence, we can extract the heat capacity as demonstrated by Hwang et al.\cite{Hwang1997}

Depicted in figure~\ref{fig:fig03_Tsingy}\,(a) we can see the full setup: The calorimeter itself sits on the front face of a cylindrical copper rod. The chip is placed on three small stems made of photo resist (see side view in figure \ref{fig:fig03_Tsingy}\,(b)) with a marginal thermal contact. This allows us to define the thermal contact, individually for each sample under investigation, by the superconducting aluminum bonding wires (green) and a $~\SI{225}{\nano\meter}$ thick, normal conducting, sputtered Au filament. A passive low-pass filter with a cut-off-frequency of about \SI{70}{\mega\hertz} is placed between the calorimeter and the dc-SQUID to suppress potential high-frequency crosstalk, in particular back action from the SQUID radiated at Josephson frequencies \cite{Roukes1999,Schwab2000a}, and to provide additional heat-sinking. The dc-SQUID itself is operated in voltage bias allowing for a typical power dissipation of less than \SI{50}{\pico\watt}. All wires are guided in epoxy glue (Stycast 2850, Henkel) to plugs. The uppermost \SI{35}{\milli\meter} of the copper rod are covered by a hollow cylinder (not shown) made of superconducting niobium to keep external magnetic stray fields, if any, constant. The SQUID array used for the two-stage readout scheme is placed in another shielded housing close-by.

\section{\label{sec:04_Results}Results and Performance}
\subsection{Temperature Sensor}

\begin{figure*}[t]
	\includegraphics[width=\textwidth]{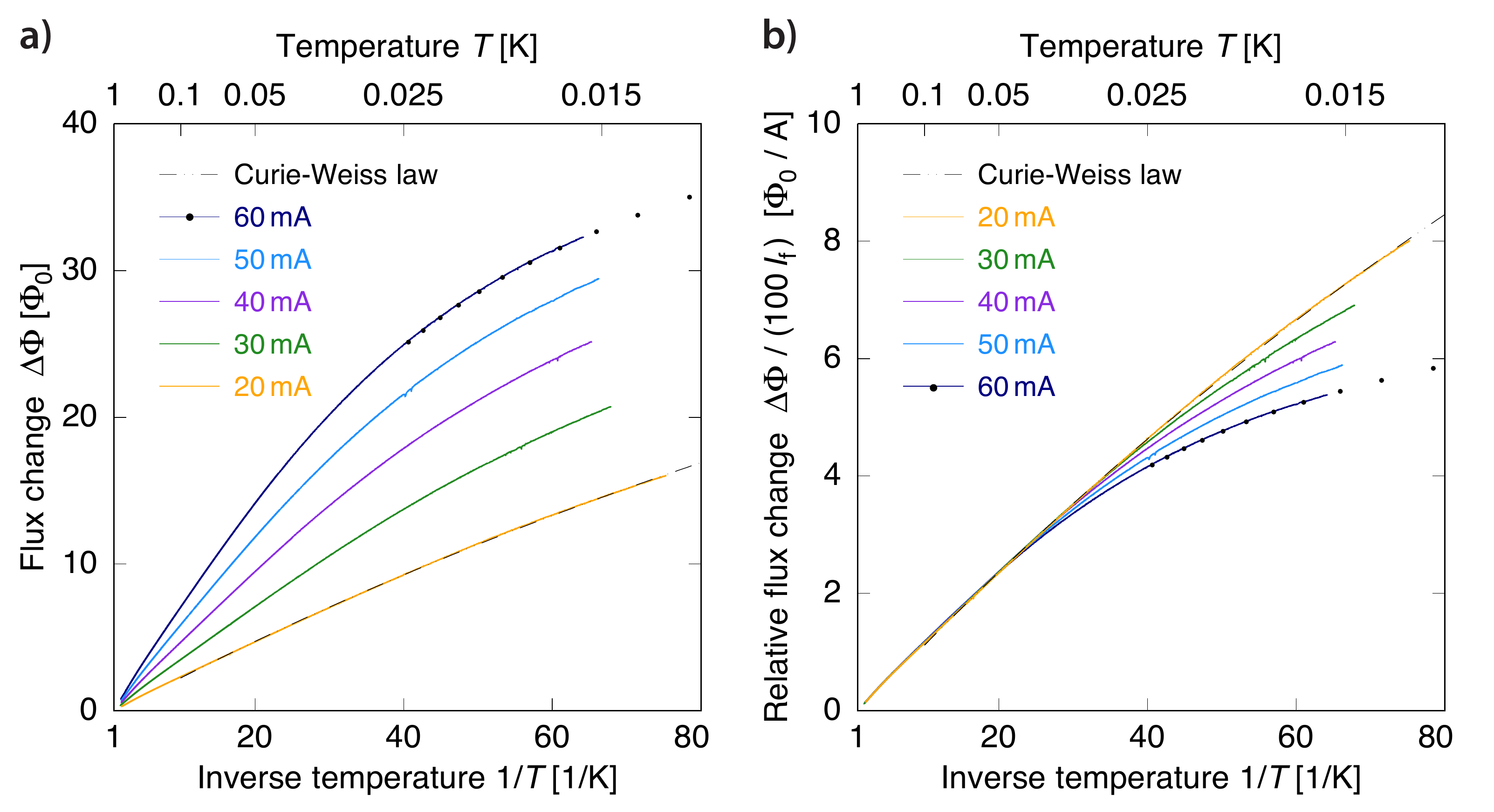}
	\caption{\label{fig:fig04_magnetization}(a) Magnetic flux change in the dc-SQUID plotted against the inverse temperature $1/T$ for various persistent currents $I_{\rm f}$. Solid lines have been measured while continuously ramping the temperature from high to low values. Additionally, single data points for $I_{\rm f} = \SI{60}{\milli\ampere}$ have been measured at stabilized, constant temperatures. (b) The graph on the right-hand side depicts the same set of data, normalized by the persistent current $I_{\rm f}$.}
\end{figure*}

In a first step, we determine the temperature dependent change of magnetization. As discussed earlier, this change of magnetization causes an observable flux change $\Delta \Phi$ in the dc-SQUID\footnote{SQUID PTB\_C633\_I03\_C6X1NM by Physikalisch-Technische Bundesanstalt, Abbestr. 2, 10587 Berlin, Germany with input inductance $L_{\rm in} = \SI{1.6}{\nano\henry}$ and  mutual input inductance $1/M_{\rm in} = \SI{5.5}{\micro\ampere/\phinought}$}. Numerical simulations indicate that the change of magnetization in the meander itself can be calculated roughly by $\Delta M = 8.8\,\textrm{A/m} \cdot \Delta\Phi / \phinought$, where $\phinought$ symbolizes the magnetic flux quantum. Figure~\ref{fig:fig04_magnetization} shows the results for five different field generating persistent currents $I_{\rm f}$. The graph on the left-hand side shows the temperature dependent flux change plotted against the inverse temperature $1/T$. The colored solid lines represent data recorded during a continuous cooldown of the experimental platform, while the black dots represent measurements performed at stabilized bath temperatures. According to the Curie law, we expect a $\Delta \Phi \propto B_{\rm 0}/T$ behavior for temperatures above \SI{50}{\milli\kelvin}. The graph on the right-hand side depicts the same set of data, normalized by the persistent current $I_{\rm f}$. In accordance with the Curie law all curves lie on top of each other at higher temperatures. At lower temperatures and higher magnetic fields, interaction and saturation effects have to be taken into account as more and more spin systems occupy the ground state. The Curie-Weiss law $\Delta \Phi \propto 1/(T+\Theta)$ can be used to describe the data in good approximation in the limit of low magnetic fields with $\Theta \approx \SI{3.3}{\milli\kelvin}$ (dashed line). Moreover, it is necessary to account for dipole-dipole and RKKY-interactions \cite{Ruderman1954,Kasuya1956,Yosida1957}, which both change the low-temperature and high-field behavior significantly. Such simulations have been performed in the case auf \underline{Au}:Er and can be adapted for \underline{Ag}:Er  yielding excellent agreement with the data \cite{Fleischmann2005,Hengstler2017}.

To underline the excellent performance of this thermometer, we evaluate the curve at $T = \SI{50}{\milli\kelvin}$ yielding $\partial\,\Delta\Phi/\partial T = \SI{-225}{\phinought/\kelvin}$. With a feedback resistance of $\SI{100}{\kilo\ohm}$, a mutual inductance of the feedback loop of $1/M_{\rm f} = \SI{12.6}{\micro\ampere/\phinought}$ and a experimentally verified readout precision $\Delta U = \SI{0,3}{\milli\volt}$ (with an averaging interval of \SI{20}{\milli\second}), we obtain a temperature resolution of 
\begin{equation}
\Delta T = \left| \frac{\partial \, \Delta \Phi}{\partial T} \right|^{-1} \frac{M_{\rm f}}{R_{\rm f}} \, \Delta U \approx \SI{0.3}{\micro\kelvin} \enskip .
\end{equation} 
This is equivalent to a temperature resolution of \SI{30}{\nano\kelvin/\sqrthz}, which becomes even better at lower temperatures due to the increasing slope of the magnetization curves. With the detector SQUID being operated in voltage bias and the low-pass filter chip between the calorimeter chip and the dc-SQUID blocking back action and serving as a heat sink, we do not observe any parasitic heating of the heat capacity chip down to below \SI{10}{\milli\kelvin}. Such heating would result in a bending-off of the magnetic flux change for lowest temperatures even in lowest fields $B_{\rm 0}$ and hence a deviation from the Curie-Weiss law.

In contrast to experiments based on suspended wires or tubes, we did not observe relevant noise contributions or heating due to mechanical vibrations. The three posts on which the calorimeter chip is placed are therefore one of the very features and novelties of the setup presented here.

\subsection{Pulse Shape Analysis}
With the temperature dependence of the magnetic flux change $\Delta \Phi$ having been calibrated, we are ready to discuss the recorded temperature response following a defined heat pulse. We used a high purity silver sample (6N, \SI{16}{\milli\gram}) placed on the electroplated gold layer of the heat capacity chip for this characterization measurement. A very thin ($\approx \SI{0.1}{\milli\meter}$) layer of vacuum grease was used as an adhesive in order to simulate the worst case of samples where we cannot make use of the metallic link at the gold sample mounting area (fig.~\ref{fig:fig02_HeatCapChip} I).

\begin{figure}
	\includegraphics[width=250pt]{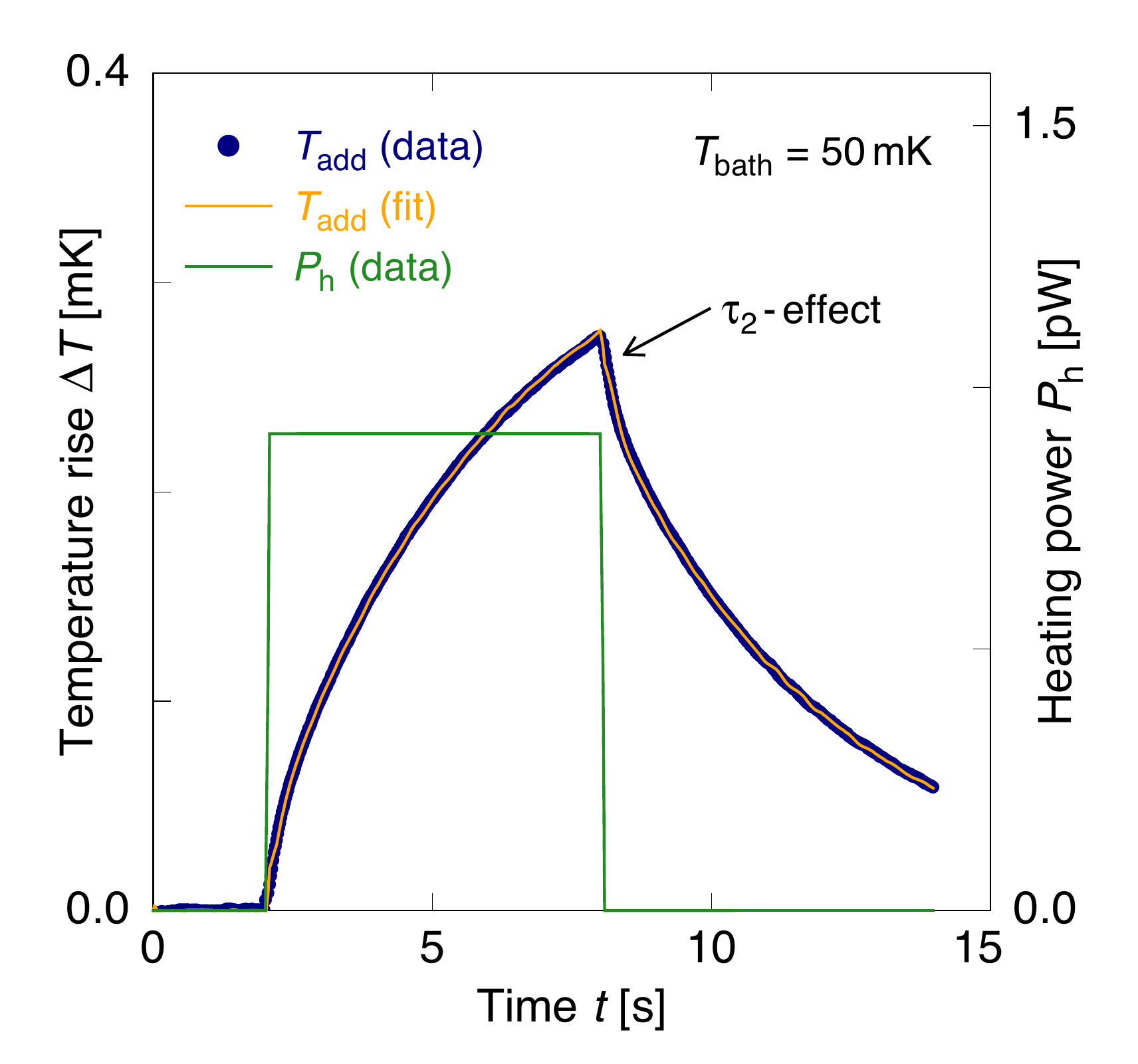}
	\caption{\label{fig:fig05_RecordedPulse}Time evolution of the heat capacity chip temperature (blue, left ordinate) when applying a well-defined heat pulse (green, right ordinate). A fit to the data following the algorithm of Hwang et al. \cite{Hwang1997} is displayed as an orange solid line.}
\end{figure}

Figure~\ref{fig:fig05_RecordedPulse} serves as an example where a typical heat pulse recorded at a bath temperature of $T = \SI{50}{\milli\kelvin}$ is displayed: A rectangular shaped heat pulse (green data, right ordinate) is applied to the AuPd-heater (figure~\ref{fig:fig02_HeatCapChip} G). As a consequence, the temperature of the addenda $T_{\rm add}$ (blue, left ordinate) increases gradually. After switching off the heating current, a relaxation towards bath temperature is monitored. We observe a somewhat faster change of the addenda temperature $T_{\rm add}$ both at the beginning and after the end of the application of the heating current $I_{\rm h}$, which is known as the $\tau_{\rm 2}$-effect: Due to the finite thermal contact between the chip and the sample, the change of the sample temperature (which is not accessible in the experiment) lags behind that of the chip itself. Due to the small addenda heat capacity, a steep temperature change $T_{\rm add}$ is observed until a sufficiently large temperature gradient between chip and sample allows for an adequate heat flow into the sample. We should add that the magnitude of the $\tau_{\rm 2}$-effect depends on an interplay between various parameters: bath temperature, thermal conductivities involved, and especially the addenda heat capacity. We found that a time resolution of \SI{20}{\milli\second} between individual data points was sufficient to model the effect adequately in our experiments even at lowest temperatures where the effect is strongest. The fit result according to the model of Hwang \textit{et al.} (orange  solid line in figure~\ref{fig:fig05_RecordedPulse}) allows us to extract the thermal conductivities $\kappa_{\rm 1}$, $\kappa_{\rm 2}$, and the heat capacity $C_{\rm tot}$.

At this point, we would like to stress once more some outstanding features of the experimental setup: First, we note again the excellent temperature resolution allowing us to resolve very small temperature changes (in the example shown, the heat pulse leads to a temperature increase of around \SI{0.5}{\percent}), while at the same time there are no noticeable temperature fluctuation due to external influences like vibrations. Furthermore, we observed no dependence of the resulting sample heat capacity $C_{\rm s}$ when varying the pulse length over one order of magnitude while keeping the absolute temperature rise constant. \\
In addition, we observed a linear increase of the extracted sample heat capacity $C_{\rm s}$ with increasing pulse height $\Delta T_{\rm max}$. This is in very good agreement with the expectation, as an increased pulse height $\Delta T_{\rm max}$ leads to a linear increase of the average sample temperature  $T + \Delta T_{\rm max}/2$ and hence a linear increase in the measured heat capacity $C_{\rm s}$ for the investigated silver sample, which 
is dominated by the contribution of the conduction electrons leading to a linearly rising heat capacity  $C_{\rm Ag} \propto T$ in the temperature range under investigation.

\subsection{Thermal Contact}

\begin{figure}
	\includegraphics[width=0.5\textwidth]{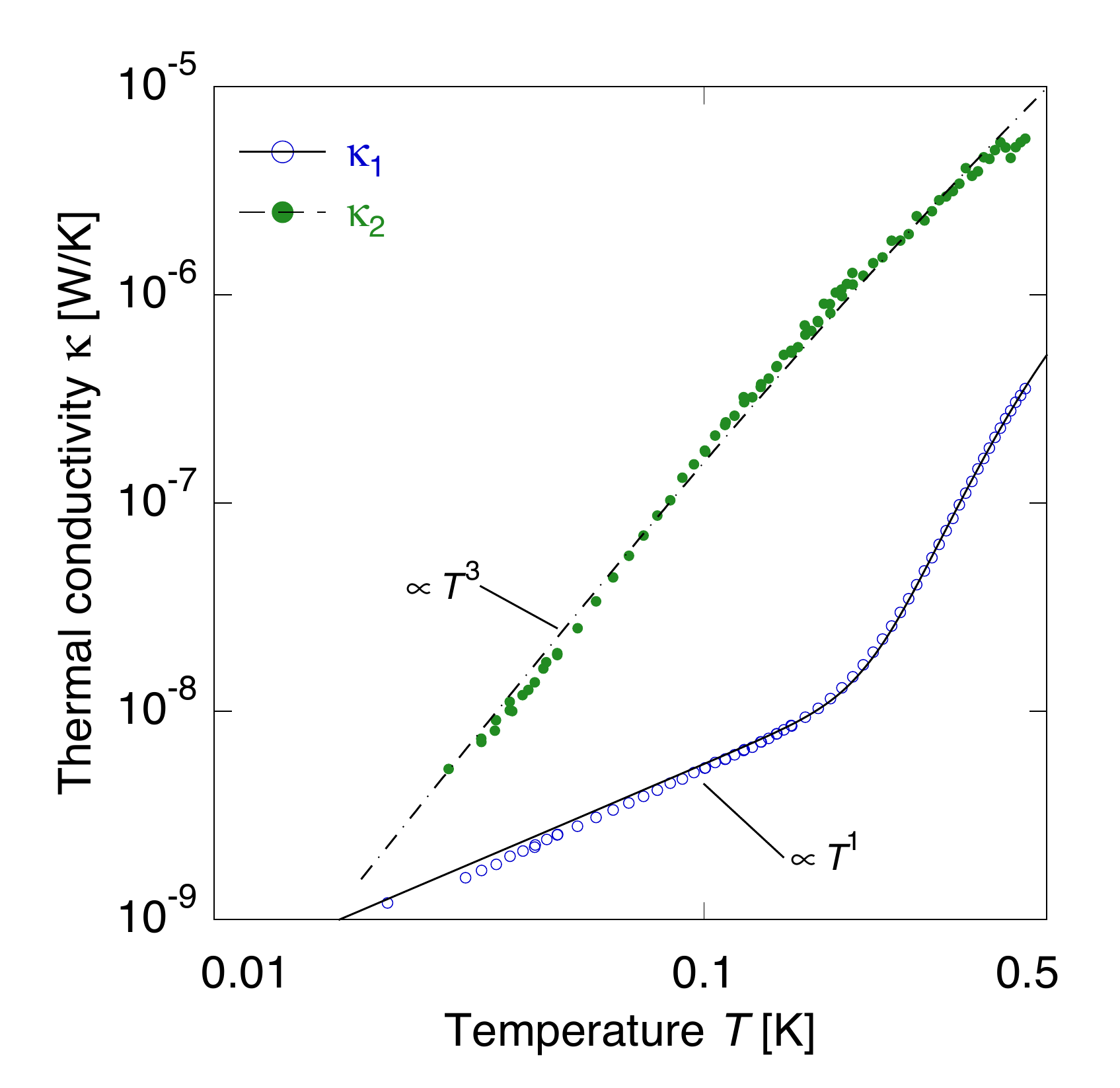}
	\caption{\label{fig:fig05_kappa}Thermal conductance $\kappa_{\rm 1}$ between chip and bath and $\kappa_{\rm 2}$ between chip and sample for a phononic contact via grease. The latter is nearly proportional to $T^3$, which is expected considering the sample's thermal boundary resistance}
\end{figure}

Another important aspect for the performance of the experimental design is the understanding and adjustment of the heat flow onto and from the chip. By fitting the full temperature response of a heat pulse, we obtain the relevant parameters, namely the thermal conductance $\kappa_{\rm 1}$ between heat capacity chip and thermal bath, and the thermal conductance $\kappa_{\rm 2}$ between the heat capacity chip and the actual sample. The data originate from measurements of the same silver sample as in the previous section. The results are displayed in figure~\ref{fig:fig05_kappa}. We start our discussion with the thermal coupling $\kappa_{\rm 2}$ to the sample: Here, we observe a nearly cubic temperature dependence $\kappa_{\rm 2} \propto T^3$. We can model the phononic heat flow between heat capacity chip and sample by a stack of three materials: electroplated gold, a thin layer of vacuum grease, and the silver sample itself. Adding up the thermal boundary resistances between each pair of layers, assumed to be similar for the ease of simplicity, and including the heat flow through the glass-like vacuum grease, we obtain
\begin{equation}
\frac{1}{\kappa_{\rm 2}} = \frac{2 R_{\rm K}}{A} \frac{1}{T^3} + \frac{l \, R_{\rm vc}}{A} \frac{1}{T^2} \enskip .
\end{equation}
With a contact area of $A = \SI{4}{\milli\meter\squared}$ and an estimated vacuum grease thickness of $l = \SI{0.1}{\milli\meter}$, we obtain a thermal boundary resistance of $R_{\rm K} = \SI{9.5e-3}{\kelvin\tothe{4}\meter\squared/\watt}$ and a thermal resistance $R_{\rm vc} = \SI{635}{\kelvin\tothe{3}\meter/\watt}$ within the vacuum grease in agreement with published data \cite{Anderson1970,Swartz1989,Pobell2007}. For metallic samples, thermal contact between the sample and the calorimeter can be realized by diffusion welding in future measurements allowing for a much better thermal coupling $\kappa_{\mathrm{2}}$.

Understanding the thermal contact $\kappa_{\rm 1}$ is another important aspect since it's adjustment allows to define the relaxation time $\tau_{\rm 1}$ for a given sample heat capacity. Starting from low temperatures, figure~\ref{fig:fig05_kappa} indicates a linear increase in $\kappa_{\rm 1}$ for temperatures below \SI{150}{\milli\kelvin} followed by a steep increase. We assume that the thermal conductivity is defined by the heat flow through the gold and aluminum bonding wires introduced in chapter~\ref{sec:03_Experimentals}, leading to 
\begin{equation}
\kappa_{\rm 1} = \frac{\mathcal{L}}{R} T + \kappa_{\rm Al,0} e^{b\,(1-T_{\rm c}/T)} \enskip ,
\end{equation}
where the first term connects the thermal conductivity of the normal-conducting gold bonding wire with its electric resistance $R$ via the Wiedemann-Franz law using the Lorenz number $\mathcal{L} = \SI{2.44}{\watt\ohm/\kelvin\squared}$. The second term denotes the thermal conductance of superconducting aluminum. By fitting the data (solid line), we obtain $R= \SI{0,48}{\ohm}$ in very good agreement with room temperatures measurements taking into account $\mathrm{RRR}\approx2$ for the sputtered gold filement (fig. \ref{fig:fig03_Tsingy} m). Furthermore, fitting values for $\kappa_{\rm Al,0} = \SI{3,7}{\micro\watt\per\kelvin}$ and $b=\num{1,5}$ are in good agreement with literature data (see\,\,\cite{Woodcraft2005} and references therein), if we assume the 10 aluminium bonding wires to have a radius of \SI{25}{\micro\meter} and an average length of \SI{5}{\milli\meter}. We would like to point out that an additional term proportional to $T^2$ or $T^3$ does not improve the quality of the fit justifying the asumption that there is no noticeable phononic heat flow through the three posts mechanically supporting the chip.

\subsection{Addenda Heat Capacity}
The last crucial property for the performance of the experimental setup we want to discuss here is the addenda heat capacity. As discussed in chapter~\ref{sec:02_Challanges}, we fabricated the micro-structured calorimeter with only small amounts of normal-conducting materials to keep the addenda heat capacity at a very low level.

\begin{figure}
	\includegraphics[width=0.5\textwidth]{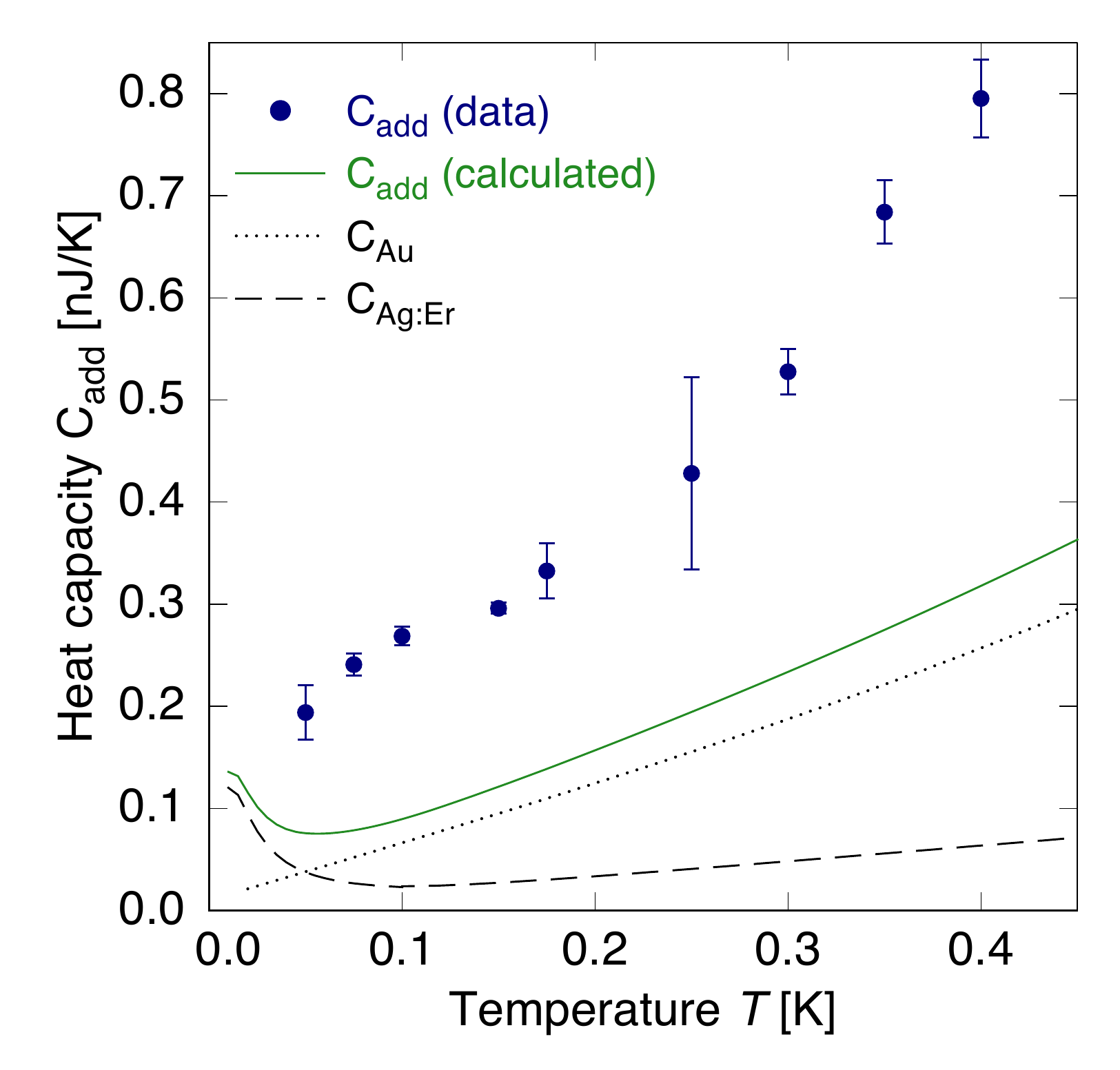}
	\caption{\label{fig:fig08_AddendaHeatCap}The main contribution to the addenda heat capacity (green solid line) is given by the sum $C_{\rm Au}+C_{\rm Ag:Er}$ of the heat capacity contributions of the electroplated gold area (dotted line) and the paramagnetic temperature sensor (dashed line). Deviations of the measured data (blue symbols) from the expected values are discussed in the text.}
\end{figure}

In fact, we expect significant contributions only from the normal-conducting Au layer as well as the Schottky contributions from the sensor material \underline{Ag}:Er. The gold sample placement area consists of a \SI{100}{\nano\meter} thick sputtered Au layer followed by \SI{500}{\nano\meter} of electroplated gold, yielding a weight of \SI{170}{\micro\gram} in total. The phononic and electronic heat capacity contributions for gold are well known\cite{Martin1973}; in addition we took into account an additional constant contribution of $5\,\mathrm{J/(m^3\,K)}$ for the sputtered part of the Au film originating from defects in the material\cite{Fleischmann2009}. The resulting contribution $C_{\rm Au}(T)$ is depicted as dotted line in figure~\ref{fig:fig08_AddendaHeatCap}.

For the sensor material \underline{Ag}:Er (\SI{14.2}{\micro\gram}), we have to take into account not only the heat capacity of silver\cite{Martin1973}, but also the contributions due to the Zeemann splitting of the 4f magnetic moments of the Er ions in the presence of the magnetic field $B_{\rm 0}$. These contributions have been calculated elsewhere for \underline{Au}:Er and should be very similar in our case \cite{Fleischmann2005}. The contribution of the sensor material $C_{\rm Ag:Er}$ for $I_{\rm f} = \SI{50}{\milli\ampere}$ is depicted as dashed line in figure~\ref{fig:fig08_AddendaHeatCap}. The resulting overall addenda heat capacity $C_{\rm add}$ is shown in green (solid line) in the same figure. The blue symbols represent measured data. First of all, we note that we observe a very small addenda of less than \SI{500}{\pico\joule/\kelvin} for temperatures below \SI{300}{\milli\kelvin}. However, we observe a roughly 2-fold higher addenda heat capacity compared to the calculated values discussed above. We were able to exclude additional contributions within the niobium layer (by varying the respective layer thicknesses) and the wafer substrate itself (sapphire- and silicon-wafers were produced). We believe that the additional contributions originate from a combination of magnetic impurities, dangling bonds, and two-level-systems\cite{Anderson1972,Phillips1972} - mainly in the amorphous SiO$_{2}$ sputtered on top of the superconducting meander-shaped coils for galvanic separation and mechanical protection. In fact, we have strong evidence from previous experiments with sputtered SiO$_{\rm 2}$ layers produced in the same sputtering unit, where we observed and proofed that magnetic impurities can significantly enhance the specific heat of those sputtered SiO$_{\rm 2}$ layers. Therefore we believe that an optimization of the sputtering procedure can further decrease the addenda specific heat in future production batches.


\section{Summary}
In this paper, we have presented a new experimental setup to measure heat capacities of mg-sized samples by means of the relaxation method down to the lowest temperatures accessible with dilution refrigerators. The microstructured heat capacity chip features a metallic, paramagnetic temperature sensor (\underline{Ag}:Er), which can be read out by a two-stage SQUID setup allowing for a temperature resolution of \SI{30}{\nano\kelvin/\sqrthz}. The voltage-biased readout of the dc-SQUID reduces its heat load to less than \SI{50}{\pico\watt} and no thermal decoupling of the heat capacity chip itself could be noted down to below $\SI{10}{\milli\kelvin}$. The electroplated gold sample placement area allows for a good thermal contact between sample and addenda. Both phononic and electronic couplings to the addenda can be realized if appropriate samples are under investigation. The thermal coupling to the bath can be adjusted according to the sample heat capacity so that relaxation times can be kept in an experimentally comfortable range. We observed an addenda heat capacity of less than \SI{500}{\pico\joule/\kelvin} for temperatures below \SI{300}{\milli\kelvin} allowing for the investigation of materials that intrinsically show a very small heat capacity such as superconductors.

\section{Data Availability Statement}
The data that support the findings of this study are openly available in a Zenodo repository at http://doi.org/10.5281/zenodo.3676882.

\section{Acknowledgements}
Part of this research was performed in the framework of the DFG project En299/5-1. The research leading to these results has received funding from the European Union’s Horizon 2020 Research and Innovation Programme, under Grant Agreement no 824109. We would like to thank Thomas Wolf for help in the preparation of the samples and Marius Hempel for support in setting up the control software of the experiment. Furthermore, we acknowledge the help of Christian Schötz, Matthäus Krantz, and the cleanroom team at the Kirchhoff Institute for Physics for technical support during device fabrication. We thank Matthew Herbst for proofreading the manuscript.

\nocite{*}
\bibliography{custom_lib}

\end{document}